%Paper: gr-qc/9302033
%From: TORRE@CC.USU.EDU
%Date: 23 Feb 1993 16:53:10 -0600 (MDT)

\input jnl
\input reforder
\ignoreuncited
\def\refto#1{[#1]}
\citeall\refto
\doublespace
\preprintno{FTG-114-USU}\dateline
\def\ss{\scriptscriptstyle}
\def\cqg{\journal Class. Quantum Grav., }
\def\ann{\journal Ann. Phys., }
\def\grg{\journal Gen. Rel. Grav., }
\title Symmetries of the Einstein Equations
\author C. G. Torre
\affil Department of Physics
Utah State University
Logan, UT  84322-4415
USA
\author I. M. Anderson
\affil Department of Mathematics and Statistics
Utah State University
Logan, UT 84322-3900
USA
\abstract\doublespace
Generalized symmetries of the Einstein equations are infinitesimal
transformations of the
spacetime metric that formally map solutions of the Einstein equations to
other
solutions.  The infinitesimal generators of these symmetries are assumed to
be local, \ie at a given spacetime point they are functions of the
metric and an arbitrary but finite number of
derivatives of the metric at the point.  We classify all generalized
symmetries of the vacuum Einstein equations in four spacetime dimensions
and find that the only generalized symmetry transformations consist of:

(i) constant scalings of the metric

(ii) the infinitesimal action of generalized spacetime diffeomorphisms.

\noindent
Our results
rule out a large class of possible ``observables'' for the gravitational field,
and suggest that the vacuum Einstein
equations are not integrable.
\endtitlepage

{\noindent\bf Introduction}

A point symmetry of a system of differential equations is a one-parameter
group of transformations of the underlying space of independent and
dependent variables that carries solutions of the equations to other
solutions.  Over a century ago, Lie \refto{Lie1896} initiated a geometrical
approach
to the study of differential equations based on their point symmetries.
By considering infinitesimal group transformations, Lie produced
algorithms for finding the point symmetries of any system of equations.
For differential equations derived from a variational principle, Noether
\refto{Noether1918}
proved that those point symmetries which preserve the action lead to
conservation laws.  Noether also pointed out that not all conservation laws
arise as a consequence of point symmetries.  She therefore introduced the
idea of derivative-dependent infinitesimal symmetry transformations, now
known as
``generalized symmetries''.  Her work, together with the appropriate
technical assumptions \refto{Olver1986},
establishes a one-to-one correspondence between generalized
symmetries of the underlying action functional and conservation laws.

In recent
years, symmetry analysis has become
an important tool in the study of differential equations \refto{Olver1986,
Bluman1989, Ovsiannikov1982}.  This is due, in part, to the intimate
connection between
generalized symmetries and integrable systems of partial
differential equations.
Indeed, a widely acknowledged attribute of an integrable field theory is the
existence of an infinite set of generalized symmetries \refto{Fokas1987,
Mikhailov1991}.   Physically, the importance of symmetries of field
equations
stems from their use in: classifying solutions to the equations,
construction of
solution generating algorithms, and, via Noether's theorem, the
identification of conservation laws.

While applied mathematicians have devoted a large amount of attention to
applications of the theory of symmetry groups to a variety of non-linear
partial
differential equations, relatively few results have been obtained for
the most important non-linear field equations of theoretical physics, \eg the
Yang-Mills equations and the Einstein equations.  The purpose of this letter
is to
report on the results of a generalized symmetry analysis of the vacuum
Einstein equations in four spacetime dimensions.  Our analysis has yielded
a complete classification of the generalized symmetries.

Why look for generalized symmetries of the Einstein equations?  The
existence of
``hidden symmetries'' of the Einstein equations would lead to solution
generating/classification techniques, and perhaps even information about
the ``general
solution'' to the Einstein equations.  There are hints that such symmetries
may exist: the two Killing field reduction of the Einstein equations leads to
an
integrable system of partial differential equations \refto{Belinsky1979,
Hauser1981}; the self-dual Einstein equations exhibit an infinite number of
symmetries and can be integrated using twistor methods
\refto{Penrose1976, Winternitz1989}.  Recently, a set of generalized
symmetries has been presented for the Einstein equations using the
Newman-Penrose formalism \refto{Gurses1993}.  A complete
generalized symmetry analysis provides a systematic and rigorous way to
unravel some aspects of the integrable behavior of the gravitational field
equations.  In particular, such an analysis indicates whether the rich
structure of special reductions of the Einstein equations extends to the full
theory, as well as clarifying the status of the symmetries of
\refto{Gurses1993}.
An equally important consequence of a generalized symmetry analysis
stems from the fact that the existence of generalized symmetries of the
Einstein
equations is a necessary condition for the existence of local differential
conservation laws for
the gravitational field.  If such symmetries/conservation laws could be
found, they would
lead to ``observables'' for the gravitational field.  It has long been an open
problem in relativity theory to exhibit such observables.  The lack of
observables currently hampers progress in canonical quantization of
general relativity \refto{Smolin1990}.

{\noindent\bf Generalized Symmetries of the Einstein Equations}

A {\it generalized symmetry} of the Einstein equations $G_{ab}=0$ is an
infinitesimal transformation $\delta g_{ab}$ of the metric which formally
maps
solutions of the Einstein equations to other ``near by'' solutions.  The
generator of a generalized symmetry transformation
is built from the spacetime position $x$, the metric, and an arbitrary but
{\it finite} number of derivatives of the metric at $x$:
$$
\delta g_{ab} = h_{ab}(x,g,\partial_x g,...) .\tag1
$$
We say that the functions $h_{ab}$ generate a symmetry if and only if they
satisfy the
linearized Einstein equations,
$$\left(-g^{dc}\delta^{(a}_{m}\delta^{b)}_n -
g^{ab}\delta^{(d}_{m}\delta^{c)}_n + 2
g^{db}\delta^{(c}_{m}\delta^{a)}_n\right) \nabla_d\nabla_c h_{ab} =
0,\tag5
$$
when the Einstein equations $G_{ab}=0$ hold.  In \(5) $\nabla_a$ is the
unique connection compatible with the Einstein metric $g_{ab}$.

	Note that while the infinitesimal transformation is a local function of
the metric, the finite transformation it generates (if it exists) can be quite
non-local.  This
is because the one-parameter family of solutions $g_{ab}(\lambda)$
generated by
$h_{ab}$ is obtained by solving a system of partial differential equations
$$
{\partial g_{ab}\over\partial\lambda}=h_{ab}( x,g,\partial_x g,...).\tag
$$

There are two classes of generalized symmetries that can be identified
immediately.  The first is the well-known scale symmetry
of the Einstein equations, which corresponds to the infinitesimal point
symmetry
$$
\delta g_{ab}=cg_{ab},\tag1a
$$
where $c$ is a constant.
The second symmetry stems from the general
covariance (diffeomorphism covariance) of the Einstein equations. It is
well-known that, for each vector field $V^a(x)$, the tensor
$$
\delta g_{ab}=\nabla_aV_b+\nabla_bV_a\tag1b
$$
satisfies the linearized equations \(5) when $G_{ab}=0$.  This point
symmetry reflects the general covariance of the Einstein equations.  The
symmetries \(1a), \(1b) comprise all the point symmetries of the vacuum
Einstein equations \refto{Ibragimov1985}.  Given a  generalized
(covariant) vector field $X_a=X_a(x,g,\partial_x g,...),$ a direct
computation shows that
$$
\delta g_{ab}=\nabla_aX_b+\nabla_bX_a\tag2
$$
also satisfies the linearized Einstein equations when $G_{ab}=0$, and
is hence a generalized symmetry.  Let us call the symmetry \(2) a {\it
generalized diffeomorphism symmetry}.   According to the principles of
general relativity, one should really view the Einstein equations as a set of
partial differential equations that determine diffeomorphism equivalence
classes of metrics.
Thus the generalized diffeomorphism symmetry \(2) is considered
physically trivial.

Our symmetry analysis shows that the scale symmetry \(1a) and
generalized diffeomorphism symmetry \(2) are in fact the {\it only}
generalized symmetries
admitted by the vacuum Einstein equations.   The proof of this fact is
rather long, accordingly it is best to begin
by classifying an important sub-class of symmetries, the ``natural''
generalized symmetries.  Natural symmetries are generated by those
$h_{ab}$ which transform properly under spacetime diffeomorphisms.
Specifically, a ``natural tensor'' $h_{ab}$ is a
tensor which is built from the metric, the curvature, and covariant
derivatives of the
curvature up to some finite order \refto{Thomas1934, Epstein1975,
Anderson1984}.  Such $h_{ab}$ are universal geometric expressions, and
are defined on any manifold irrespective of its topological structure.     In
this case we have the following
theorem.

\noindent {\bf Theorem 1:}  {\it Let $\delta g_{ab} = h_{ab}(g,\partial_x
g,...)$
be a natural generalized symmetry for the Einstein equations $G_{ab}=0$
in four spacetime dimensions.
Then
$$
h_{ab} = c g_{ab} + 2 \nabla_{(a}X_{b)},\tag3
$$
where $c$ is a constant and
$$
X_a=X_a(g,\partial_x g,...)\tag4
$$
is a natural (covariant) vector field.}

We will now sketch the proof of this theorem; a more rigorous discussion
with all of the details will be given elsewhere \refto{Bigpaper}.  All our
considerations are
local in spacetime, \ie we do not worry about boundary conditions, so
our results are quite general.  The strategy is to view $h_{ab}$ as a
collection of functions on a very large space
${\cal J}^N$ parametrized by the spacetime point $x$, the metric, the
curvature, and the first
$N$ covariant derivatives of the curvature at $x$.  Note that ${\cal J}^N$
is {\it finite} dimensional.  For example, ${\cal J}^0$ is the 34-dimensional
space parametrized by $(x, g_{ab}, R_{abcd})$, where $R_{abcd}$ is the
curvature tensor.
The Einstein equations $G_{ab}=0$, and their derivatives
$\nabla_cG_{ab}=0,\nabla_c\nabla_dG_{ab}=0,\dots$, define a subspace
${\cal E}^N$ of ${\cal J}^N$
and it is only on this subspace that the equations \(5) need to be satisfied.
In the example above, ${\cal E}^0$ is a 24-dimensional submanifold of
${\cal J}^0$ defined by the equations
$$
g^{ac} R_{abcd}=0.\tag
$$
Using the chain rule, the linearized Einstein equations \(5)
become an overdetermined system of partial differential equations
restricting the functional form of $h_{ab}$.  Because the
resulting system of equations is quite overdetermined, it is
possible to find all of its solutions.

The primary
complication in the analysis is that \(5) need only be satisfied on ${\cal
E}^N$.
To handle this complication we appeal to work of Penrose
\refto{Penrose1960, Penrose1984} which
gives coordinates for ${\cal E}^N$ in terms of {\it spinors}.  The
introduction of two-component spinors at this point restricts our analysis to
the conventional four-dimensional spacetime of
Einstein's general theory of relativity.  While it is quite possible that our
results generalize to higher dimensions, as it stands our analysis is
limited to the four-dimensional case.

Penrose's ``exact set of fields'' for the vacuum Einstein equations
\refto{Penrose1960, Penrose1984} can be
viewed as giving an explicit parametrization of ${\cal E}^N$.  Let
$\Psi_{\ss
ABCD}$ represent the Weyl spinor (capital Latin indices are
two-component spinor indices).  Penrose's result is that the symmetrized
covariant derivatives of the Weyl spinor
$$\Psi_{\ss A_1\cdots A_{n+4}}^{\ss A_1^\prime\cdots A_n^\prime}
:= \nabla_{\ss (A_1}^{\ss (A_1^\prime}\cdots\nabla_{\ss A_n}^{\ss
A_n^\prime)}\Psi_{\ss A_{n+1}A_{n+2}A_{n+3}A_{n+4})}\tag11
$$
and its complex conjugate, for $n=0,1,\dots,$ are freely specifiable at a
point
of an Einstein space and completely determine the curvature and all
covariant derivatives of the
curvature at that point.  We will denote the spinor \(11) and its complex
conjugate by
$\Psi^n$ and $\overline\Psi^n$ respectively.  Using the spinor form of the
Ricci and Bianchi identities on an Einstein space it is straightforward to
show that $\Psi^n$ satisfies
$$
\nabla_{\ss B}^{\ss B^\prime}\Psi_{\ss A_1\cdots A_{n+4}}^{\ss
A_1^\prime\cdots A_n^\prime}=\Psi_{\ss A_1\cdots A_{n+4}B}^{\ss
A_1^\prime\cdots A_n^\prime B^\prime} +  {\cal Q}\tag
$$
where ${\cal Q}$ denotes terms involving $\Psi^k$ and $\overline\Psi^k$
for $k=0,\dots,n-1$.

The spinor translation of
the generalized symmetry equation \(5) can be put into the form:
$$\eqalign{
\Bigg[-\epsilon^{\ss AB}\epsilon^{\ss A^\prime B^\prime}\epsilon_{\ss
MR} \epsilon_{\ss NS}\delta^{\ss M^\prime}_{\ss R^\prime} \delta^{\ss
N^\prime}_{\ss S^\prime}
&+\delta^{\ss A}_{\ss M}\epsilon^{\ss A^\prime
N^\prime}\Big(\epsilon_{\ss NS}\delta^{\ss
B}_{\ss R} \delta^{\ss B^\prime}_{\ss R^\prime}\delta^{\ss
M^\prime}_{\ss S^\prime}
+ \epsilon_{\ss NR} \delta^{\ss
B}_{\ss S}\delta^{\ss B^\prime}_{\ss S^\prime}\delta^{\ss M^\prime}_{\ss
R^\prime}
\Big)\Bigg] \cr
&\times \nabla_{\ss AA^\prime}\nabla_{\ss BB^\prime}h^{\ss
MN}_{\ss M^\prime N^\prime}=0,}\tag10
$$
where $h^{\ss MN}_{\ss M^\prime N^\prime}$ is the spinor form of the
symmetry generator $h_{ab}$.
We have also introduced the skew-symmetric $\epsilon$-spinors, which are
used to raise and lower spinor indices.  Equation \(10) is to be satisfied
modulo the Einstein equations.  Hence, without loss of generality, we can
assume in
\(10) that $h^{\ss
MN}_{\ss M^\prime N^\prime}$ is a function of the soldering form
$\sigma_a^{\ss AA^\prime}$, where
$$
g_{ab} = \sigma_a^{\ss AA^\prime}\sigma_{b{\ss AA^\prime}},\tag
$$
and the $\Psi^n, \overline\Psi^n$-spinors up to some finite derivative order
$N$.
The natural generalized symmetry $h^{\ss MN}_{\ss M^\prime
N^\prime}$ must satisfy \(10) for all values of the
$\Psi^n$ spinors and their complex conjugates.

As an illustration of our analysis let us assume that $N=1$ so that $h_{ab}$
is a natural
tensor depending on no more than three derivatives of the metric at a given
point:
$$
h_{ab}=h_{ab}( \sigma,\Psi^0,\overline\Psi^0,\Psi^1,\overline\Psi^1).\tag
$$
Because \(10) involves two derivatives of $h_{ab}$, the left-hand side of
\(10) is a function of the soldering form and $\Psi^n$, $\overline\Psi^n$-
spinors
  for $n=0,\dots, 3$.  Differentiating \(10) with respect to $\Psi^3$ leads to
the
following restriction on the dependence of $h_{ab}$ on the $\Psi^1$-
spinors:
$$h^{\ss M(NABCDE)}_{\ss M^\prime N^\prime E^\prime} + h^{\ss
M(NABCDE)}_{\ss E^\prime N^\prime M^\prime}=0,\tag12$$
where we have defined
$$h^{\ss MNABCDE}_{\ss M^\prime N^\prime E^\prime}
:={\partial h^{\ss MN}_{\ss M^\prime N^\prime}\over\partial
\Psi_{\ss ABCDE}^{\ss E^\prime}}.\tag$$
Differentiating \(10) with respect to $\overline\Psi^3$ shows that the
complex conjugate of \(12) holds also.

Next, let us demand that the second derivative of $\(10)$ with respect to
$\Psi^2$ vanish for all values of the
$\Psi^n$-spinors.  Using \(12) this leads, after some analysis, to
$$
{\partial^2 h^{\ss MN}_{\ss M^\prime N^\prime}\over\partial \Psi_{\ss
ABCDE}^{\ss E^\prime}\partial\Psi_{\ss
RSTUV}^{\ss V^\prime}} =0.\tag
$$
Thus $h_{ab}$ is {\it linear} in its dependence on $\Psi^1$.  Similar
computations involving second derivatives of \(10) with respect to $\Psi^2,
\overline\Psi^2$ and $\overline\Psi^2, \overline\Psi^2$ show that
the symmetry must be linear in its dependence on the
$\overline\Psi^1$-spinors, and also that $h_{ab}$ contains no terms
involving products $\overline\Psi^1\Psi^1$.

We have thus found that the spinor expression of the generalized symmetry
$h_{ab}$
takes the form
$$
h^{\ss MN}_{\ss M^\prime N^\prime}=
A_{\ss M^\prime N^\prime E^\prime}^{\ss MNABCDE}
\Psi_{\ss ABCDE}^{\ss E^\prime} + B^{\ss MN}_{\ss M^\prime
N^\prime} + {\rm complex\ conjugate\hfill},\tag
$$
where the $A$ and $B$ spinors depend on the soldering form and the
undifferentiated
Weyl spinors $\Psi^0, \overline\Psi^0$.  The condition \(12) gives us
further information about the spinor $A$;
\(12) is satisfied if and only if there exist spinors $D$ and $X$ such that
$$A^{\ss MNABCDE}_{\ss M^\prime N^\prime E^\prime}=
\epsilon^{\ss M(A} D^{\ss CDE}_{\ss M^\prime N^\prime
E^\prime}\epsilon^{\ss B)N}  +
X_{\ss M^\prime}^{\ss M(BCDE}\epsilon^{\ss A)N}\delta^{\ss
N^\prime}_{\ss E^\prime} +
X_{\ss N^\prime}^{\ss N(BCDE}\epsilon^{\ss A)M}\delta^{\ss
M^\prime}_{\ss E^\prime}.\tag13
$$

We further restrict the structure of the $A$-spinor by taking the mixed
second partial derivative of \(10) with respect to $\Psi^2$ and $\Psi^1$.
After considerable analysis, the
resulting equations can be shown to imply: (i) the $D$-spinor in \(13) is
independent of
$\Psi^0$ and $\overline\Psi^0$, \ie $D$ is a function of the soldering form
only; and (ii) the spinor $X$ is
a gradient with respect to $\Psi^0$, \ie
$$
X_{\ss M^\prime}^{\ss MBCDE} =
{\partial X^{\ss M}_{\ss M^\prime}\over\partial \Psi_{\ss BCDE}}.\tag14
$$
Here $X_ {\ss M}^{\ss M^\prime}$ is the spinor form of a natural
spacetime vector field.

Given \(14), the $X$-spinors correspond to a generalized diffeomorphism
symmetry \(2).  This can be seen by comparing the coefficient of the
$\Psi^1$ term in the spinor form of \(2), namely
$$
\nabla_{\ss A}^{\ss A^\prime}X_{\ss B}^{\ss B^\prime} + \nabla_{\ss
B}^{\ss B^\prime}X_{\ss A}^{\ss A^\prime} = {\partial X_{\ss B}^{\ss
B^\prime}\over\partial \Psi_{\ss RSTU}}\Psi_{\ss RSTUA}^{\ss
A^\prime} + {\partial X_{\ss A}^{\ss A^\prime}\over\partial \Psi_{\ss
RSTU}}\Psi_{\ss RSTUB}^{\ss B^\prime} + {\rm complex\
conjugate},\tag
$$
with the last two terms of \(13).

To summarize, we have found that the only generalized symmetry for
$N=1$ is a linear function of $\Psi^1$.  Modulo terms of the form \(2), the
coefficient $A$ of the $\Psi^1$ term is a natural spinor built from the
soldering
form only.  However, it can be shown that there is no spinor with
the symmetries of $A$ that is built solely from the soldering form.
Therefore, modulo generalized diffeomorphism symmetries, the linear
term in $\Psi^1$ vanishes and we
conclude that (modulo \(2)) the symmetry can only depend on $\Psi^0,
\overline\Psi^0$ and the soldering form:
$$
h_{ab}= 2 \nabla_{(a}X_{b)} +
h^\prime_{ab}(\sigma,\Psi^0,\overline\Psi^0).\tag
$$

We now repeat the whole analysis starting with \(10) and ending with \(14)
under the
assumption that the symmetry only depends on the soldering form and the
undifferentiated Weyl spinors. A virtually identical series of calculations
proves
that $h^\prime_{ab}$ is a function of the
soldering form only, $h^\prime_{ab}=h^\prime_{ab}(\sigma)$.  This is
easily seen to imply that the point symmetry $h^\prime_{ab}$ can only be
the scaling
symmetry \(1a), \ie $h^\prime_{ab}=cg_{ab}$.  This proves the theorem
when $N=1$.

The proof of Theorem 1 in general is again via induction in the dependence
of the symmetry $h_{ab}$ on derivatives of the metric. The spinor
equations that arise in the analysis for $N>1$ are considerably more
complicated than in the example above, but they can be solved using
elementary spinor techniques.

Theorem 1 still leaves open the possibility that a non-trivial generalized
symmetry can be constructed in some non-universal way from the metric,
say with the aid of some
externally prescribed auxiliary fields.  Such symmetries are not universal
in the sense that their existence may depend on the underlying manifold
structure of the spacetime.  While it is hard to know {\it a priori} what the
geometrical
meaning of such a symmetry could be, if we view the Einstein equations as
just some set of partial differential equations, and ignore their geometric
structure, then it is natural to seek such
symmetries.  We do this by allowing the
symmetry generator $h_{ab}$ to be an explicit function of the spacetime
position $x$ and drop
the assumption that it is built only from the metric, curvature and
covariant derivatives of curvature.   Thus $h_{ab}$ is now allowed to
depend on the
spacetime position, the metric and a finite number of derivatives of the
metric at
a point, with no tensoriality assumptions.  It is possible to generalize the
analysis that leads to Theorem 1 to this case, and we can prove the more
general theorem \refto{Bigpaper}:

\noindent
{\bf Theorem 2 :}  {\it Let $\delta g_{ab} = h_{ab}(x,g,\partial_x g,...)$
be a
generalized symmetry for the Einstein equations $G_{ab}=0$ in four
spacetime dimensions.  Then
$$
h_{ab} = c g_{ab} + 2 \nabla_{(a}X_{b)}
$$
where $c$ is a constant and
$$
X_a=X_a(x,g,\partial_x g,...).
$$}

The proof of Theorem 2 involves the enlargement of the spinor variables
to include the non-tensorial parts of the metric derivatives.  An inductive
proof, similar to that used in the natural case, is used to reduce the
derivative dependence of the symmetry generator to only first derivatives
of the metric modulo terms of the form \(2).  Because of the complicated
dependence of the linearized Einstein equations on first and second
derivatives of the metric occurring in the covariant derivatives, this case---
first-order generalized symmetries---must be treated separately.  A lengthy
analysis leads to the result that, modulo the generalized diffeomorphism
symmetry, the symmetry generator is a function of the undifferentiated
metric only.  This leads back to the scale symmetry and completes the
proof.

Theorems 1 and 2 allow us to determine the structure of possible
conservation laws for the vacuum Einstein equations.  Let us define a {\it
local differential conservation law} for the Einstein equations as a current
(vector density) $J^a=J^a(x,g,\partial_x g,\ldots)$ satisfying
$$
\nabla_a J^a = \partial_a J^a = 0, \tag8
$$
when $G_{ab}=0$.  Then we can prove the following corollary.

\noindent
{\bf Corollary 1:}  {\it Let $J^a$ be a local differential conservation law
for the vacuum Einstein equations in four spacetime dimensions.  Then
there exists functions $S^{ab}(x, g, \partial_x g, \cdots)$ skew symmetric
on $a$ and $b$, $S^{ab}=-S^{ba}$, such that, up to terms that vanish when
the Einstein equations hold,
$$
J^a = \nabla_b S^{ab}.\tag9
$$
If  $J^a$ is a natural vector density, \ie built from the metric, curvature
and covariant derivatives of the curvature, then $S^{ab}$ can be chosen to
be a natural tensor density.}

This corollary follows from the fact that the existence of a generalized
symmetry is a necessary condition for the existence of a local differential
conservation law (see, \eg \refto{Olver1986}), and some fundamental
results from the theory of the variational bicomplex
\refto{Anderson1992}.  From the point of view of the theory of local
differential conservation laws, the form \(9) of $J^a$ is trivial in the sense
that such conservation laws are {\it always} possible for any system of
equations, irrespective of the form of the equations.  Nevertheless, such
currents can have a physical role to play in general relativity.   Indeed, \(9)
forms the basis for constructing energy-momentum pseudo-tensors for the
gravitational field \refto{Goldberg1980}.

The form \(9) of the conservation laws
has strong implications for the existence of ``observables''
in the Hamiltonian formulation of gravitation in closed universes.  Recall
that observables are
defined as functions on the gravitational phase space that have weakly
vanishing Poisson brackets with the super-Hamiltonian and
super-momentum \refto{Sundermeyer1982, Karel}.  This is
equivalent to defining observables as constants of motion for the
Einstein equations.  From \(9), however, it is clear that if the spatial
manifold is compact without boundary there can be no non-trivial constants
of motion built as spatial integrals of local functions of the spacetime
metric and its derivatives. Thus our generalized symmetry analysis has
ruled out a large class of observables.  In particular, we conjecture that
there are no
observables built as spatial integrals of local functions of the canonical
(ADM)
coordinates and momenta (and their derivatives).  It would thus appear that
observables must be constructed in a non-local fashion, \eg
along the lines of those found in the class of cylindrically symmetric
Einstein metrics in \refto{CGT1991c}.  We are currently exploring the
Hamiltonian implications of our analysis.

{\noindent\bf Summary}

We have classified all generalized symmetries and local differential
conservation laws of the vacuum Einstein equations in four spacetime
dimensions.  The symmetries consist of constant scalings and the induced
action of infinitesimal generalized diffeomorphisms.  The corresponding
conservation laws are trivial.  We note that the vacuum Einstein equations,
when viewed as a system of equations for diffeomorphism equivalence
classes of metrics,
fail to pass a widely acknowledged ``litmus test'' for the integrability of a
system of partial differential equations,
namely, the existence of an infinite-dimensional set of generalized
symmetries \refto{Fokas1987, Mikhailov1991}.

Our analysis suggests several questions for further study; they include: the
Hamiltonian interpretation of Theorems 1, 2 and Corollary 1, existence of
generalized symmetries of subsystems of the Einstein equations, existence
of non-local symmetries, \eg Backlund transformations, and existence of
generalized symmetries of the Einstein equations with matter couplings.

\noindent{\bf Acknowledgments}

This material is based upon work supported by the National Science
Foundation under Grant No. DMS-9100674 (IMA), and a Utah State
University Faculty Research Grant (CGT).

\references\doublespace
\refis{Bigpaper}{I. M. Anderson and C.G. Torre, in preparation 1993.}

\refis{Karel}{Kucha\v r has shown there are no non-trivial observables
that are linear functionals of the ADM momenta; K. V. Kucha\v r, \jmp 22,
2640, 1981.}

\refis{Gurses1993}{M. Gurses, \prl 70, 367, 1993.}

\refis{ADM1959}{R. Arnowitt, S. Deser, and C. Misner, \pr 116, 1322,
1959.}

\refis{ADM1962}{R. Arnowitt, S. Deser, and C. Misner in {\it
Gravitation: An
Introduction to Current Research}, edited by L. Witten (Wiley, New York
1962).}

\refis{Anderson1984}{I. M. Anderson, \journal Ann. Math., 120, 329,
1984.}

\refis{Anderson1992}{I. M. Anderson, ``Introduction to the Variational
Bicomplex'', in {\it Mathematical Aspects of Classical Field Theory} (Eds.
M. Gotay, J. Marsden, V. Moncrief),\journal Cont. Math., 32, 51, 1992.}

\refis{Arms1979}{J. Arms, \jmp 20, 443, 1979.}

\refis{Arms1980}{J. Arms, \jmp 21, 15, 1980.}

\refis{Ashtekar1991a}{A. Ashtekar, {\it Lectures on Non-Perturbative
Canonical
Gravity}, (World Scientific, Singapore 1991) and references therein.}

\refis{Ashtekar1991b}{ A. Ashtekar, L. Bombelli, and O.
Reula, in {\it Mechanics,
Analysis and Geometry : 200 Years After Lagrange}, edited by M.
Francaviglia ( North-Holland, New York 1991). }

\refis{Barrow1982}{J. Barrow, \prpts 85, 1, 1982.}

\refis{Belinsky1979}{V. Belinsky and V. Zakharov, \journal Sov. Phys.
JETP, 50, 1, 1979.}

\refis{Bluman1989}{G. Bluman and S. Kumei, {\it Symmetries of
Differential Equations}, (Springer-Verlag, New York 1989).}

\refis{Choquet1982}{Y. Choquet-Bruhat, C. DeWitt-Morette, and M.
Dillard-Bleick, {\it Analysis, Manifolds and Physics}, (North-Holland,
New York 1982). }

\refis{Dirac1964}{P.A.M. Dirac, {\it Lectures on Quantum Mechanics},
(Yeshiva
University, New York, 1964).}

\refis{Epstein1975}{D. Epstein, \journal J. Diff. Geom., 10, 631, 1975.}

\refis{Fischer1979}{See A. Fischer and J. Marsden in {\it General
Relativity: An
Einstein Centenary Survey}, edited by S. Hawking and W. Israel
(Cambridge University Press, Cambridge 1979).}

\refis{Fokas1987}{A. Fokas, \journal Stud. Appl. Math., 77, 253, 1987.}

\refis{Goldberg1980}{J. Goldberg in {\it General
Relativity
and Gravitation: 100 Years After the Birth of Albert Einstein, Vol. 1},
edited by A. Held (Plenum, New York 1980).}

\refis{Halliwell1991}{J. Halliwell, \prd 43, 2590, 1991.}

\refis{Hauser1981}{I. Hauser and F. Ernst, \jmp 22, 1051, 1981.}

\refis{Henneaux1991}{G. Barnich, M. Henneaux, C. Schomblond, \prd 44,
R939, 1991.}

\refis{Hormander1966}{L. H\" ormander, \journal Ann. Math., 83, 129,
1966.}

\refis{Ibragimov1985}{N. Ibragimov, {\it Transformation Groups
Applied to Mathematical Physics}, (D. Reidel, Boston 1985).}

\refis{Isenberg1982}{J. Isenberg and J.
Marsden, \prpts 89, 181, 1982, and references therein.}

\refis{Isham1985}{C. J. Isham and K. V. Kucha\v r, \ann 164, 288, 1985;
\ann 164, 316, 1985.}

\refis{Kuchar1971}{K. V. Kucha\v r, \prd 4, 955,
1971.}

\refis{Kuchar1972}{K. V. Kucha\v r, \jmp 13, 758, 1972}.

\refis{Kuchar1976}{K. V. Kucha\v r, \jmp 17, 801, 1976.}

\refis{Kuchar1978}{K. V. Kucha\v r, \jmp 19, 390, 1978.}

\refis{Kuchar1981}{K. V. Kucha\v r, \jmp 22, 2640, 1981.}

\refis{Kuchar1992}{K. V. Kucha\v r, ``Time and Interpretations of
Quantum Gravity'', to
appear in the proceedings of {\it The Fourth Canadian Conference on
General Relativity and Relativistic Astrophysics}, edited by G. Kunstatter,
D. Vincent, and J. Williams (World Scientific, Singapore 1992).}

\refis{Lanczos1970}{C. Lanczos, {\it The Variational Principles of
Mechanics} (University of Toronto Press, Toronto 1970).}

\refis{Lie1896}{S. Lie, {\it Geometrie der Beruhrungstransformationen},
(B. G. Teubner, Leipzig 1896).}

\refis{Mikhailov1991}{A. Mikhailov, A. Shabat, and V. Sokolov in {\it
What is Integrability?}, ed. V. Zakharov (Springer-Verlag, New York
1991).}

\refis{Noether1918} {E. Noether, \journal Nachr. Konig. Gesell. Wissen.
Gottinger Math. Phys. Kl., , 235, 1918.}

\refis{Olver1986}{P. Olver, {\it Applications of Lie Groups to
Differential Equations}, (Springer-Verlag, New York 1986).}

\refis{Osgood}{See, {\it Conceptual Problems
of Quantum Gravity}, edited by A. Ashtekar and J. Stachel, (Birkh\"auser,
Boston 1991).}

\refis{Ovsiannikov1982}{L. Ovsiannikov, {\it Group Analysis of
Differential Equations}, (Academic Press, New York 1982).}

\refis{Penrose1960}{R. Penrose, \ann 10, 171, 1960.}

\refis{Penrose1976}{R. Penrose, \grg 7, 31, 1976.}

\refis{Penrose1984}{R. Penrose and W. Rindler, {\it Spinors and Space-
Time, Vol. 1}, (Cambridge University Press, Cambridge 1984).}

\refis{Pullin1991}{J. Pullin in {\it Relativity and Gravitation: Classical and
Quantum}, edited by J. C. D'Olivo {\it et al}, (World Scientific, Singapore
1991).}

\refis{Smolin1990}{C. Rovelli and L. Smolin, \np B331, 80, 1990.}

\refis{Sundermeyer1982}{K. Sundermeyer, {\it Constrained Dynamics},
(Springer-Verlag, Berlin 1982).}

\refis{Teitelboim1976}{A. Hanson, T. Regge, C. Teitelboim, {\it
Constrained Hamiltonian Systems}, (Accademia Nazionale dei Lincei,
Rome 1976).}

\refis{Thomas1934}{T. Y. Thomas, {\it Differential Invariants of
Generalized Spaces}, (Cambridge University Press, Cambridge 1934).}

\refis{CGT1989}{K. V. Kucha\v r, C. G. Torre, \jmp 30, 1769, 1989.}

\refis{CGT1990}{K. V. Kucha\v r and C. G. Torre, \prd 43, 419, 1990.}

\refis{CGT1991a}{K. V. Kucha\v r and C. G. Torre, \prd 44,
3116, 1991.}

\refis{CGT1991b}{K. V. Kucha\v r and C. G. Torre in {\it Conceptual
Problems
of Quantum Gravity}, edited by A. Ashtekar and J. Stachel, (Birkh\"auser,
Boston 1991).}

\refis{CGT1991c}{ C. G. Torre, \cqg 8, 1895, 1991.}

\refis{CGT1992a}{C. G. Torre, ``Is General Relativity an `Already
Parametrized' Theory?'', Utah State University Preprint, 1992.}

\refis{CGT1992b}{C. G. Torre, ``Covariant Phase Space Formulation of
Parametrized Field Theories'', Utah State University Preprint, 1992.}

\refis{CGT1992c}{C. G. Torre, in preparation, 1992.}

\refis{Wald1990}{J. Lee and R. Wald, \jmp 31, 725, 1990.}

\refis{Winternitz1989}{C. Boyer and P. Winternitz, \jmp 30, 1081, 1989.}

\refis{Witten1987}{C. Crnkovic and E. Witten in {\it 300 Years of
Gravitation},
edited by S. Hawking and W. Israel (Cambridge University Press,
Cambridge 1987).}

\refis{York1980}{Y. Choquet-Bruhat and J. York in {\it General
Relativity
and Gravitation: 100 Years After the Birth of Albert Einstein, Vol. 1},
edited by A. Held (Plenum, NY 1980).}

\endreferences
\endit